\documentclass{article}
\usepackage{setspace}
\usepackage{epsfig}
\begin{document}
\title{\bf Continuum Electrostatics in Cell Biology}
\author{L. John Gagliardi\\
Rutgers-the State University\\
Camden, New Jersey 08102\\
e-mail: gagliard@camden.rutgers.edu}
\maketitle
\begin{abstract}
Recent experiments revealing possible nanoscale
electrostatic interactions in 
force generation at kinetochores for chromosome 
motions have prompted 
speculation regarding possible models for interactions 
between positively charged molecules in 
kinetochores and negative charge on 
C-termini near the plus ends of microtubules. A clear picture of
how kinetochores establish and maintain a dynamic coupling to
microtubules for force generation during the
complex motions of mitosis remains elusive.
The current paradigm of molecular cell biology requires that specific
molecules, or molecular geometries, for force generation be identified.
However it is possible to account for mitotic motions within a classical
electrostatics approach in terms of experimentally known cellular
electric charge interacting over nanometer distances. These charges are
modeled as bound surface and volume continuum charge distributions. 
Electrostatic consequences of intracellular pH changes 
during mitosis may provide a master clock for the events of mitosis.
\end{abstract}
\noindent 
{\it Keywords}: Electrostatics; Mitosis; Chromosome motility; 
Intracellular pH  
%\hspace{0.2in} {\it Keywords}: Electrostatics; Mitosis; Chromosome 
%motility 
%\newpage
%\doublespacing
%
\section{Introduction} 
\label{sec:level1}
Although the molecular biology paradigm has been 
quite successful in 
solving a number of problems in the understanding of cell structure and 
function, important problems remain open.  Some of the most notable of 
these are in the area of cell division.  Given the constraints wittingly 
or unwittingly imposed by the molecular cell biology paradigm, it would 
seem that models of mitotic motions and events have become more and more 
complex, 
and therefore -- to this observer at least -- more and more unsatisfactory. 
%In any event, 
Insisting on a molecular biology approach to explaining 
mitotic chromosome motions (and a number of other cellular processes) 
%can be likened to the ancient Greeks' insisting on 
has striking similarities to the ancient Greeks' insisting on 
perfect circles to explain planetary motions. 
This paper reviews an alternative based on 
classical electrostatics expressed in terms of 
stably bound continuum surface and volume charge densities ({\it charge 
distributions}). This sort of approach %offers the possibility of 
makes it possible to describe the dynamics, including timing and 
sequencing, of  post-attachment mitotic motions 
within a comprehensive approach. 
\newline \newline \noindent
The known charge in mitotic chromosome motions is the %net 
negative charge on chromosome arms and centrosomes, and positive 
charge at kinetochores. 
Negative charge at and near {\it plus} ends of microtubules and positive 
charge at {\it minus} ends of microtubules will be assumed. (According to
existing convention, one end is designated {\it plus} because  
of its more rapid growth, there being no reference to charge in the use
of this nomenclature.)
Arguments for these two assumptions will be presented; however, they 
are to be viewed here as the sole postulates within %primarily viewed as 
which a comprehensive model for post-attachment mitotic movements and 
events can be framed. 
\newline \newline \noindent
In 2002 and 2005 papers, I argued that indirect 
experimental evidence indicates that pole-facing plates of kinetochores 
manifest positive charge 
[1-3] and interact with negatively charged 
microtubule free ends to provide the motive force for poleward force 
generation at kinetochores. This has subsequently been supported 
by experiments [4-6] implicating positively charged molecules at %IN  
kinetochores in establishing a dynamic coupling to negative 
charge on microtubules during mitosis. %%% replace REF 4 %%%%%%
\newline \newline \noindent
Assuming a volume positive charge at kinetochores and negative charge 
at and near the free plus ends of 
microtubules, it was possible to derive a magnitude of the maximum 
(tension) force per microtubule for poleward chromosome motions that falls 
within the experimental range [3]. In these papers, 
I also proposed that indirect %%% "indirect" 
experimental evidence is consistent with a negative charge distribution 
on centrosomes [1-3]. 
Recent direct experimental measurements have confirmed this [7].  
\newline \newline \noindent 
As noted above, a major advantage of focusing on cellular charge 
distributions is 
that it appears to offer the possibility of discovering a minimal 
assumptions %comprehensive 
model for post-attachment chromosome motions [3,8]. 
A model of this sort can point the way to the eventual discovery of specific 
molecules, and their biochemistries, that are responsible for the various 
mitotic motions. This is what is now happening. As mentioned above, 
a number of recent experiments have shown that certain kinetochore 
molecules that bind with 
microtubules have a net positive charge, and that poleward force for 
chromosome motions at kinetochores may be due to electrostatic 
interactions between these 
molecules and negative charge on microtubules [4-6]. 
These discoveries might 
have been made sooner if the above mentioned 2002 and 2005 papers 
had been duly noted. %and referenced. %%%DISCUSS HERE POSSIBLE REASONS: 
\section{Some cellular electrostatics} 
In the cytoplasmic medium (cytosol) within 
biological cells, it has been generally thought that electrostatic 
fields are subject to 
strong attenuation by screening with oppositely charged ions 
(counterion screening), decreasing exponentially to much 
smaller values over a distance of several {\it Debye lengths}. 
The Debye length within cells is typically given to be of order 1 nm 
[9], and since cells of interest in the present 
work (i.e. eukaryotic) can be taken to have much larger dimensions, 
one would be tempted to conclude 
that electrostatic force could not be a major factor in providing the 
cause for mitotic chromosome movements in biological cells. 
However, the presence of 
microtubules, as well as other factors to be discussed shortly, 
change the picture completely. 
\newline \newline \noindent
Microtubules can 
be thought of as intermediaries that extend the reach of the 
electrostatic interaction over cellular distances, making the 
second most potent force in the universe available to cells 
in spite of their ionic nature. 
%
%\begin{figure}[htbp] \centering
%\includegraphics[scale=.30]{microt2.eps}
%\caption{A microtubule highlighting a protofilament. A B-lattice 
%microtubule}   
%\end{figure}
%
\noindent
Microtubules are 25 nm diameter cylindrical structures comprised 
of {\it protofilaments}, each consisting of tubulin 
dimer subunits, 8 nm in length, aligned lengthwise parallel to the 
microtubule axis. The protofilaments are bound laterally to form a 
sheet that closes to form a cylindrical microtubule. 
The structure of microtubules is similar in all eukaryotic cells. Cross 
sections reveal that the wall of a microtubule consists of a circle of 
4 to 5 nm diameter subunits. The circle typically contains 13 subunits 
as observed in vivo. 
Neighboring dimers along protofilaments exhibit a small (B-lattice) 
offset of 0.92 nm from protofilament to protofilament. 
\newline \newline \noindent
Microtubules continually assemble and disassemble, 
so the turnover of tubulin is ongoing. 
The characteristics of microtubule lengthening (polymerization) and
shortening (depolymerization) follow a pattern known as ``dynamic
instability": that is, at any given instant some of the microtubules
are growing, while others are undergoing rapid breakdown.
In general, the rate at which microtubules undergo net assembly -- or
disassembly -- varies with mitotic stage [10]. 
Changes in microtubule dynamics are integral to changes in the 
motions of chromosomes during the stages of mitosis. 
%
%\newline \newline \noindent
%
Poleward and antipoleward chromosome motions occur 
%intermittently 
during prometaphase and metaphase. Antipoleward  motions dominate during 
the {\it congressional} movement of chromosomes to the cell {\it 
equator}, and poleward motion prevails during anaphase-A.   
It is assumed here that poleward chromosome motions are in response to 
disassembling kinetochore microtubules at kinrtochores and poles, 
and antipoleward chromosome motions 
are in response to assembling microtubules at chromosome arms.  
%Assembling and disassembling microtubules are depicted in Figure 2. 
%
%\begin{figure}[htbp] \centering   
%\includegraphics[scale=.30]{dynamic2.eps}
%\caption{Shrinking (showing protofilament curling) and growing 
%microtubules}
%\end{figure}
%%
\noindent
Experiments have shown that the intracellular pH (${\rm pH_i}$) 
of many cells rises to a maximum at the onset of mitosis, subsequently
falling during later stages [11,12]. 
Studies [13] %[Schatten {\it et al.}, 1985] 
have shown that {\it in vivo} 
microtubule growth (polymerization) is favored by higher pH values. 
It should be noted that {\it in vitro} studies of the role of pH in
regulating microtubule assembly indicate a pH optimum for 
assembly in the range of 6.3 to 6.4. The disagreement between {\it in 
vitro} and {\it in vivo} studies %regarding microtubule polymerization 
has been analyzed in relation to the nucleation potential of microtubule 
organizing centers like centrosomes [13], %[Schatten {\it et al.}, 1985], 
and it has been suggested that ${\rm pH_i}$ regulates the nucleation 
potential of microtubule organizing centers [14-16]. %[Kirschner, 1980; 
%De Brabander {\it et al.}, 1982; Deery and Brinkley, 1983]. 
This favors the more complex physiology characteristic of {\it in vivo}
studies to resolve this question. 
\newline \newline \noindent 
Kinetochore molecules are known to self-assemble onto condensed, 
negatively charged 
DNA at centromeres [17], %[Alberts {\it et al.}, 1994b], 
indicating that 
kinetochores may exhibit positive charge at their pole-facing plates. 
This is an example of an important aspect of electrostatic 
interactions within cells: namely their longer range compared to 
other intracellular molecular 
interactions, and the resulting capacity of electrostatic 
forces to organize molecules and structures within cells. 
This line of reasoning was the basis for my assuming 
positive charge on pole-facing ``plates" of kinetochores in a previous 
paper [3]. In earlier works I had 
assumed positive charge at kinetochores 
for different reasons [1,2]. 
\newline \newline \noindent
Cellular electrostatics is strongly influenced by significantly reduced 
counterion screening due to layered water adhering to 
charged molecules. Such water layering -- with 
consequent reduction or elimination of Debye screening -- to charged 
proteins has long been theorized [18,19] 
%[Jordan-Lloyd and Shore, 1938; Pauling, 1945] 
and has been confirmed by experiment [20]. 
%[Toney {\it et al.}, 1994]. 
Additionally, water between sufficiently close (up to 4 nm) 
charged proteins has a dielectric constant that 
is considerably reduced from the {\it bulk} value far from charged 
surfaces [3,8,21]. As will be discussed in the next section, 
this would further increase the tendency for 
an electrostatic assist to aster and spindle self-assembly. 
\newline \newline \noindent
The combination of these two effects (or conditions) -- water layering 
and reduced dielectric constant -- can significantly influence cellular 
electrostatics in a number of important ways. This is especially true 
in relation to mitosis [8,21].  
\newline \newline \noindent
The aster's pincushion-like appearance is consistent with 
electrostatics since electric dipole subunits will 
align radially outward about a 
central charge with the geometry of the resulting configuration 
resembling the electric field of a point charge. 
From this it seems reasonable to assume that the pericentriolar 
material, the {\it centrosome matrix} within which the microtubule  
dimer dipolar subunits assemble in many %%%% animal 
cell types to form an aster [22], %[Joshi {\it et al.},1992], 
carries a net charge. This agrees with observations that 
the microtubules appear to start in the centrosome matrix %%%% region 
[23]. %[Wolfe, 1993a]. 
One may assume that the sign of this charge is negative [1,2]. 
%[Gagliardi, 2002a; Gagliardi, 2002b].  
This assumption is consistent with experiments [24] revealing that  
%[Heald {\it et al.}, 1996] %%ALSO CONSISTENT WITH POS CHGE AT - ENDS!!
mitotic spindles can assemble around DNA-coated beads incubated 
in {\it Xenopus} egg extracts. The phosphate groups of the DNA will 
manifest a net negative charge at the pH of this experimental system. 
This experimental result was cited in my 2002 and 2005 papers to conclude 
that centrosomes are negatively charged [1-3]. 
As noted above, centrosomes have recently been shown to 
have a net negative charge by direct measurement [7]. 
%[Horme\~{n}o {\it et al.}, 2009]. 
%
\newline \newline \noindent 
A number of investigations have focused on the electrostatic 
properties of microtubule tubulin subunits [25-28]. 
%[Satari\'{c} {\it et al.}, 1993; Brown and Tuszy\'{n}ski, 
%1997; Baker {\it et al.}, 2001; Tuszy\'{n}ski {\it et al.}, 1998]. 
Large scale calculations of the tubulin molecule 
%[Tuszy\'{n}ski et al., 2002] 
have been carried out using molecular dynamics programs 
%%%{\it Tinker} [Dudek and Ponder, 1995], 
along with protein parameter sets. %{\it Charmm} [Brooks et al., 1983]. 
The dipole moment of tubulin has been calculated to be as large as 1800 
Debye (D) [25,29]. %[Brown and Tuszy\'{n}ski, 1997; 
%Tuszy\'{n}ski {\it et al.}, 1995]. 
In experiments carried out at nearly physiological conditions, 
the dipole moment has been determined to be 36 D [30], 
%[Stracke et al., 2002], 
corresponding to a dipole charge of approximately 0.1 electron 
per dimer. %%%%%%%%%%%%% CHECK THIS CALCULATION %%%%%%%%%%%%%%%%%%%%%%%
Experiments [29,31] %[Tuszy\'{n}ski {\it et al.}, 1995; Sackett, 1997] 
have shown that tubulin net charge depends strongly on pH, varying 
quite linearly  from --12 to --28 (electron charges) 
between pH 5.5 and 8.0. This could be significant for 
microtubule dynamics during mitosis because, as noted 
above, many cell types exhibit a decrease of 0.3 to 0.5 pH units from a 
peak at prophase during mitosis. 
\newline \newline \noindent
It has been determined that tubulin has a large overall negative 
charge of 20 (electron charges) at pH 7, and that as much as 40 \% 
of the charge resides on C-termini [32]. The C-termini can extend 
perpendicularly outward from the microtubule axis as a function of 
${\rm pH_i}$, extending 4--5 nm at ${\rm pH_i}$ 7 [32]. 
%[Tuszy\'{n}ski, 2002]. 
%
It would therefore seem reasonable to assume that an increased tubulin 
charge and the resulting greater extension of C-termini may be integral 
to an increased probability for microtubule 
assembly during prophase when ${\rm pH_i}$ is highest. 
%[Steinhardt and Morisawa, 1982]. 
This will be discussed next.  
\section{Intracellular pH as a clock for mitosis}
As noted above, in addition to addressing force generation 
for post-attachment chromosome motions, a continuum electrostatics 
approach to mitotic motions can also account for the 
timing and sequencing of the detailed changes in these 
motions. %%RECHECK THIS STATEMENT 
These changes can be 
attributed to changes in microtubule dynamics based on a progressively 
increasing microtubule disassembly to assembly %% DISASSEMBLY/ASSEMBLY
ratio for kinetochore microtubules that is caused by a steadily 
decreasing ${\rm pH_i}$ during mitosis [2,8]. 
\newline \newline \noindent
A higher ${\rm pH_i}$ during prophase is consistent with an enhanced 
interaction between highly extended C-termini of tubulin dimers 
and positively charged regions of neighboring dimers. 
This enhanced interaction is due to their greater extension as well 
as increased expression of negative charge, with 
both favoring microtubule growth.  
It would therefore seem reasonable to expect that prophase high 
${\rm pH_i}$ conditions %%%%%%% THE INFLUENCE OF PROPHASE .... 
and the electrostatic 
nature of tubulin dimer subunits greatly assists in their self-assembly 
into the microtubules of the asters and spindle [1,2,8]. 
As discussed in %Section 2, 
the previous section, this self-assembly would be aided by 
significantly reduced counterion screening due to layered water and the 
reduced dielectric constant between charged protein surfaces. 
An electrostatic component to the biochemistry of the microtubules 
in assembling asters is consistent with experimental observations of pH 
effects on microtubule assembly [13], as well as the sensitivity of 
%[Schatten {\it et al.}, 1985], 
microtubule stability to calcium ion concentrations [33,34]. 
%[Weisenberg, 1972; Borisy and Olmsted, 1972]. 
%
\newline \newline \noindent
The two effects (or conditions) discussed in %Section 2 
the last section would be expected to significantly 
increase the efficiency of microtubule self-assembly in asters and  
spindles by (1) allowing electrostatic interactions over % significantly 
greater distances than Debye (counter-ion) screening dictates, 
and (2) increasing the 
strength of these interactions by an order of magnitude due to a 
corresponding order of magnitude reduction in the cytosolic dielectric 
constant between charged protein 
surfaces separated by critical distances or less. 
\newline \newline \noindent 
Thus it would seem reasonable to assume that, over %%% the critical 
distances consistent with the reduced dielectric constant and 
modified counterion screening discussed above, 
the electrostatic nature of tubulin dimers would %%%will%%%may %%%could   
allow tubulin dimer microtubule subunits (1) to be attracted to and align 
around charge distributions 
within cells -- in particular, as mentioned above, around 
centrosomes -- and (2) to align end to end and laterally, facilitating 
the formation of asters and mitotic spindles [1,2,8]. 
\newline \newline \noindent
The motive force for the migration of asters and assembling spindles  
during prophase can also be understood %%%addressed 
in terms of nanoscale electrostatics.  
As a consequence of the negative charge %at and near the free plus ends 
on microtubules at, and on C-termini near, the plus free ends  
of microtubules of the forming asters and %nascent 
half-spindles, the asters/half-spindles would be %repelled 
continuously repelled 
electrostatically from each other and drift apart. Specifically, 
as microtubule assembly proceeds, a subset of the negatively charged 
microtubule free ends at and near the periphery of one of the growing 
asters/forming half-spindles would mutually repel a subset of the 
negatively charged free ends at and near the periphery of the other 
growing aster/half-spindle, causing the asters/half-spindles to drift 
apart as net assembly of microtubules continues and subsets of 
interacting microtubules are continually replaced [1,2,8]. 
Microtubules disassembling from previously overlapping configurations 
could also generate repulsive force between asters/half-spindles, but 
net microtubule assembly will dominate during prophase. 
%[Gagliardi, 2002b]. %Polymerization of dimer subunits will take place 
%
\newline \newline \noindent
As discussed above, because of significantly reduced counterion 
screening and the low 
dielectric constant of layered water adhering to charged 
tubulin dimers, the necessary attraction and alignment 
of the dimers during spindle self-assembly would be 
enhanced by the considerably increased range and strength of the 
electrostatic attraction between oppositely charged regions of 
nearest-neighbors. %nearest-neighbor dimers. 
Similarly, the 
mutually repulsive electrostatic force between subsets of %%% the 
like-charged free plus ends of interacting microtubules from opposite 
half-spindles in the growing mitotic spindle would be expected to be 
significantly increased in magnitude and range. %%%increased.
Thus mutual electrostatic repulsion of the negatively charged 
microtubule plus ends distal to centrosomes in assembling 
asters/half-spindles could provide the driving force for 
their poleward migration in the forming spindle [1,2].   
%[Gagliardi, 2002b]. %%%%%%%%%CHECK THIS%%%%%%%%%%%%%%%%%%%%%%%%%%%%
A subset of interacting microtubules in a small portion of %%%the 
a forming spindle is depicted in Figure 1. 
\newline \newline \noindent 
As noted above, it is 
important to recognize that interacting 
microtubules can result from either growing or shrinking microtubules 
but polymerization probabilities will dominate during prophase. 

\begin{figure}[htbp] \centering
\includegraphics[width = .7 \textwidth]{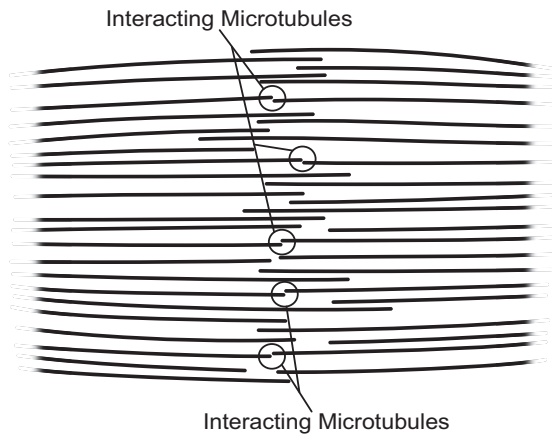}
\caption{A subset of interacting microtubules in a small portion 
of a forming mitotic spindle. Protofilament curling for disassembling 
microtubules is not shown on this scale}
\end{figure} 
\noindent
As cited above, experiments 
have shown that ${\rm pH_i}$ of many cell types 
rises to a maximum at the onset of mitosis, %during prophase 
subsequently falling steadily through mitosis. Although it is   
experimentally difficult to resolve the exact starting time for the 
beginning of the decrease in ${\rm pH_i}$ during the cell cycle, it 
appears to decrease 0.3 to 0.5 pH units from the typical peak values 
of 7.3 to 7.5 measured earlier during prophase. 
The further decrease in ${\rm pH_i}$ through metaphase [12] 
%[Steinhardt and Morisawa, 1982]
would result in increased instability of
the microtubules comprising the spindle fibers. Previously, I noted   
that {\it in vivo} experiments have shown that
microtubule stability is related to ${\rm pH_i}$, with a more basic
pH favoring microtubule assembly. 
\newline \newline \noindent
It is important to note that pH in the vicinity of
the negatively charged %exposed 
plus ends of microtubules (see discussion of net charge at microtubule 
free ends below) will be even lower %%%FIRST MENTION OF FREE END CHGE?
than the bulk ${\rm pH_i}$ because of the effect of %%% THE
negative charge at the free plus ends of the microtubules.
This lowering of pH in the vicinity of negative charge distributions
is a general result. Intracellular pH in such limited volumes is often
referred to as {\it local} pH.
As one might expect from classical Boltzmann
statistical mechanics, the hydrogen ion concentration
at a negatively charged surface can be shown to be the product of the
bulk phase concentration and the Boltzmann factor $e^{-e\zeta/kT}$,
where $e$ is the electronic charge, $\zeta$ is the (negative) electric
potential at the surface, and $k$ is Boltzmann's constant [35]. 
%[Hartley and Roe, 1940]. 
For example, for typical
mammalian cell membrane negative charge densities, and therefore typical
negative cell membrane potentials, the local pH can be reduced 0.5 to 1.0
pH unit just outside the cell membrane. Because of the negative
charge at the plus ends of microtubules, a reduction of
pH would be expected in the immediate vicinity of these free ends
making the local pH influencing microtubule dynamics considerably lower,
and a lower bulk ${\rm pH_i}$ would be accompanied by an even lower
local pH. %further increasing the tendency for net
%microtubule disassembly (as bulk values decrease through mitosis).
%
\newline \newline \noindent 
A continuum electrostatics model of mitotic events also addresses 
the dynamics of nuclear envelope fragmentation and reassembly [36]. 
Experimentally observed increases in whole cell sialic acid content [37] 
%Glick, 1971
and intracellular pH during prophase [11,12], 
followed by an observed release of 
free calcium from nuclear envelope stores at the onset of nuclear 
envelope breakdown [38,39] could
significantly enhance the manifestation of negative charge on
the nuclear envelope, providing sufficient electrostatic  
energy for nuclear envelope fragmentation [36]. 
Experimental observations regarding the mechanical properties of the
plasma membrane show that electrostatic stress does manifest itself in   
ways consistent with this scenario [40].
\newline \newline \noindent
Since terminal sialic acids are attached to membrane proteins that 
are firmly anchored in the lipid bilayer, the observed 
disassembly of the nuclear envelope is consistent with electrostatic 
repulsion between membrane continuum charge clusters, which 
tear apart under the influence of increased electrostatic charge [36]. 
It is difficult to envision a purely biochemical process that would 
result in the nuclear envelope's breaking into fragments of many 
molecules each. 
Models for nuclear envelope breakdown in the current literature do not 
address this. 
The observed lowering of both intracellular pH and whole cell sialic acid
content during late anaphase and telophase 
for many cell types [11,12,37] is 
consistent with a decreased manifestation of net negative charge on 
membrane fragments at that time. 
These decreases could shift the balance of thermal energy versus 
electrostatic repulsive energy, % in favor of thermal energy,  
allowing the closer approach of membrane fragments necessary for
reassembly to occur in nascent daughter cells [36].
\newline \newline \noindent
An increased probability for
microtubule depolymerization, as compared to the prophase predominance
of microtubule assembly, is consistent with alternating poleward
and antipoleward motions 
-- with antipoleward motions more probable -- 
of {\it monovalently} attached chromosomes during prometaphase. 
As discussed elsewhere [2,3], after a {\it bivalent} attachment to 
both poles, 
poleward forces toward both poles acting in conjunction with
inverse square antipoleward forces exerted between negatively charged 
microtubule free plus ends and negatively charged 
chromosome arms could account
for chromosome congression. The relative complexity of microtubule
disassembly force generation at kinetochores and poles coupled with
inverse square antipoleward forces from microtubule assembly at
chromosome arms precludes an unequivocal %CLEAR %%%%%%%%%%%%%%%
conclusion regarding a possible continuing increase in the microtubule
disassembly to assembly (disassembly/assembly) probability ratio during 
chromosome congression.
However, metaphase chromosome 
midcell oscillations are indirect experimental evidence 
for a microtubule disassembly/assembly probability
ratio approaching unity. 
\newline \newline \noindent 
At late metaphase,
before anaphase-A, experiments reveal that the poleward motions of sister
kinetochores stretch the intervening centromeric chromatin, producing
high kinetochore tensions. It is reasonable to attribute these high
tensions to a continuing microtubule disassembly/assembly 
probability ratio increase caused 
by a further lowering of ${\rm pH_i}$. The resulting attendant
increase in poleward electrostatic disassembly force on sister chromatids 
would lead to increased tension.
A lower ${\rm pH_i}$ would also increase the expression of positive
charge on sister kinetochores, with the possibility of 
further increasing the tension due to increased mutual repulsion. 
\newline \newline \noindent 
Thus regarding post-attachment chromosome movements through
metaphase, it seems reasonable to ascribe
an increasing microtubule dissassembly/assembly probability
ratio, with attendant changes in microtubule dynamics and
associated mitotic chromosome motions through metaphase, 
to an experimentally
observed steadily decreasing ${\rm pH_i}$. We may then envision the
decrease in ${\rm pH_i}$ from a peak at prophase favoring microtubule
assembly, declining through prometaphase as discussed above, and
continuing to decline through metaphase when  
parity between microtubule assembly and disassembly leads to
midcell chromatid pair oscillations, culminating in 
increased kinetochore disassembly tension close to anaphase-A, 
as the cell's 
master clock controlling microtubule dynamics, and consequently 
the events of mitosis. One might also be tempted to attribute 
the more complete dominance of microtubule disassembly -- with an
accompanying predominance of poleward
electrostatic disassembly forces --
during anaphase-A to a further continuation of a decreasing
intracellular pH. However, as discussed elsewhere [3,8,21], any
additional possible decreases in ${\rm pH_i}$ during anaphase-A may not
be a major determinant of anaphase-A motion. 
\section{Continuum electrostatics in mitotic force\\generation}
Following a {\it monovalent} attachment to one pole, chromosomes
are observed to move at considerably slower speeds, a few ${\rm \mu m}$
per minute, in subsequent motions throughout prometaphase [41]. 
%[Grancell and Sorger, 1998].
In particular, a period of slow motions toward and away from a pole   
will ensue, until close proximity of the negatively charged end of a
microtubule from the opposite pole with the other ({\it
sister}) kinetochore in the
chromatid pair results in an attachment to both poles (a {\it
bivalent} attachment) [2,3]. %%% GARDNER - VERBIAGE
Attachments of additional microtubules from both poles will
follow. (There may have been additional attachments to the first pole  
before any attachment to the second.) After the sister kinetochore 
becomes attached to microtubules from the opposite pole, chromosomes 
perform a slow (1--2$\,{\rm \mu m}$ per minute) congressional 
motion to the spindle equator, culminating in 
oscillatory motion of chromatid pairs during metaphase. 
\newline \newline \noindent
Chromosome motion during anaphase has two major components, designated as 
anaphase-A and anaphase-B. Anaphase-A is concerned with the poleward 
motion of chromosomes, accompanied by the shortening of kinetochore 
microtubules at kinetochores and/or spindle poles. The second component, 
anaphase-B, involves the separation of the poles. Both 
components contribute to the increased separation of chromosomes during 
mitosis. 
\newline \newline \noindent 
Molecular biology explanations of these motions require that 
specific molecules, and/or molecular geometries, for mitotic 
chromosome force generation be identified for each motion. 
As indicated above, electrostatic models 
within the molecular cell biology paradigm 
have recently been sought (or advanced) 
involving positively charged kinetochore molecules 
interacting with negative charge on microtubules. 
As in the situation involving models that center partially or wholly on 
simulations, these molecular biology approaches are quite complex and 
primarily attempt to address specific mitotic motions, most notably 
poleward force generation at kinetochores. 
Critical experimental observations such as the ``slip-clutch" 
mechanism [42], observations of calcium ion concentration on anaphase-A 
motion, %{Zhang et al., 1990] 
%[Maiato et al., 2004] %%observations of increased tension 
and polar generation of poleward force are not addressed. 
However, it is possible to account for the dynamics of post-attachment 
mitotic motions in terms of electrostatic interactions between  
experimentally known, stably bound continuum %%%%MUTUALLY INTERACTING 
surface and volume electric charge distributions 
interacting over nanometer distances. 
This is the approach that I have taken in a series of papers 
[1-3, 21] and book [8]. %%Check numbering 
\newline \newline \noindent 
As mentioned above, charge disributions are known to exist at centrosomes, 
chromosome arms, and kinetochores. Assumptions of negative charge at 
microtubule plus ends and positive charge at microtubule minus ends 
 -- the only assumptions -- are sufficient to explain the dynamics, 
timing, and sequencing of post-attachment chromosome motions. %and events.  
These assumptions will now be discussed. %%motivated. 
\newline \newline \noindent 
Excluding possible contributions from microtubule associated proteins, 
the evidence for a net negative charge at microtubule plus ends and net 
positive charge at minus ends is as follows: (1) large scale computer 
calculations of tubulin dimer subunits indicate that 18 positively charged 
calcium ions are bound within $\beta$ monomers with an equal number of 
negative charges localized at adjacent $\alpha$ monomers [25,26], 
%[Satari\'{c}, 1993; Brown, 1997], 
(2) experiments reveal that microtubule 
plus ends terminate with a crown of $\alpha$ subunits, and minus ends 
terminate with $\beta$ subunits [43], 
(3) the lower local pH vicinal to a negatively charged
centrosome matrix would cause a greater expression of positive charge at
microtubule minus ends, 
(4) the higher pH vicinal to a positively charged kinetochore pole-facing 
``plate" would cause a greater expression of 
negative charge at microtubule plus ends, 
(5) negative charge on centrosome matrices will induce positive charge on 
microtubule minus ends, and positive charge at pole-facing plates of 
kinetochores will induce negative charge on microtubule plus ends. 
\newline \newline \noindent
As discussed elsewhere [3,8,21], force generation from 
positive charge at the free minus ends of kinetochore microtubules 
may be responsible for polar generation of poleward force. A calculation 
of the force per microtubule assuming positive  
charge at microtubule minus ends falls within the experimental 
range [3]. Although a calculation of induced positive charge 
on a microtubule minus end from negative charge on a centrosome 
matrix is difficult because of the complex 
geometry, a reciprocal calculation of induced negative charge on a 
centrosome matrix from positive charge at the minus of a microtubule 
is relatively straightforward, and agrees with experimental ranges 
for cellular charge densities and force per microtubule 
measurements [21]. 
\newline \newline \noindent 
Similarly, net positive charge at kinetochore 
pole-facing surfaces would induce negative 
charge on the plus ends of kinetochore microtubules proximal to 
kinetochores, and reciprocal calculations at kinetochores similar to those 
at centrosomes (see previous paragraph) are also in agreement with 
experimental ranges for cellular charge densities and force per 
microtubule measurements [21].
\newline \newline \noindent 
The above charge distributions at plus and minus microtubule free ends 
are also in accord with the common observation that the free ends of an 
aster/half-spindle's microtubules distal to centrosomes (the pinheads 
in a pincushion analogy) are not attracted 
to the negatively charged outer surface of the nuclear envelope. 
If this were not the case, the forming half-spindles 
would not be able to move freely 
in their migration to the poles of the cell. 
\newline \newline \noindent 
As mentioned above, critical experimental observations such as the 
``slip-clutch'' mechanism and calcium ion concentration effects on 
anaphase-A motion have not been addressed by current models for mitotic 
chromosome motions. 
These experiments will now be reviewed along with their natural 
explanations within 
the context of a continuum electrostatics approach. 
\newline \newline \noindent 
At the high kinetochore tensions prior to anaphase-A mentioned above, 
coupled microtubule plus ends often switch from a depolymerization 
state to a polymerization state of dynamic instability.
This may be explained by kinetochore microtubule plus or minus 
free ends taking up the slack by polymerization to sustain 
attachment and resist further centromeric chromatin stretching. This
is known as the ``slip-clutch mechanism" [42].
%[Maiato {\it et al.}, 2004].
%
\newline \newline \noindent 
The slip-clutch mechanism is addressed within the context of the present 
work as follows: (1) microtubule assembly at a kinetochore 
or pole is regarded as operating in passive response to a repulsive 
{\it robust} inverse square electrostatic antipoleward microtubule 
assembly force acting between the plus ends of 
astral microtubules and chromosome arms [2,3] and/or an electrostatic  
microtubule disassembly force at a sister kinetochore or at poles [3]; 
(2) non-contact electrostatic forces acting over 
a range of protofilament free end distances (up to 4 nm, as discussed 
above) from bound positive charge 
 -- both inside and near ``surfaces" -- at kinetochores 
would be effective in maintaining coupling 
while larger protofilament gaps in the same or other 
microtubules are passively filled in; 
(3) the repulsive inverse square electrostatic assembly force 
acting at the sister chromatid's arms will 
provide a positive feedback mechanism to resist detachment. 
This explanation of the slip-clutch 
mechanism follows as a direct consequence of the present 
approach to chromosome motility with no additional assumptions. 
\newline \newline \noindent 
There appears to be an optimum calcium ion 
concentration for maximizing the speed of chromosome
motions during anaphase-A. If the [${\rm Ca^{2+}}$] is increased to  
micromolar levels, anaphase-A chromosome motion is increased two-fold
above the control rate; if the concentration is further
increased slightly beyond the optimimum, the chromosomes will slow
down, and possibly stop [44]. %\cite{Zhang90}.
It has long been recognized that one way elevated
[${\rm Ca^{2+}}$] could increase the speed of chromosome motion during
anaphase-A is by facilitating microtubule 
depolymerization [33,45-48], and it has 
%\cite{Weisenberg72,Salmon80,Kiehart81,Cande81,Olmsted75}, and it has 
been commonly believed that microtubule depolymerization, if not the
motor for chromosome motion, is at least the rate-determining step 
[49-52]. However, the 
%\cite{Nicklas75,Nicklas87,Salmon75,Salmon89}. However, the
slowing or stopping of chromosome motion associated with moderate
increases beyond the optimum [${\rm Ca^{2+}}$] is more difficult to
interpret since the microtubule 
network of the spindle is virtually intact and
uncompromised. Such disruption of the mitotic spindle requires much 
higher concentrations [44,53]. %\cite{Zhang90,Wolfe425}.
\newline \newline \noindent
In terms of the present model, higher concentrations of
doubly-charged calcium ions would shield the negative charge at      
the {\it plus} ends of kinetochore microtubules as well as negative 
charge at the 
centrosome matrix, shutting down the poleward-directed nanoscale
electrostatic disassembly force. 
\newline \newline \noindent 
An experimental test of nonspecific divalent cation effects on
anaphase-A chromosome motion by substitution of
${\rm Mg^{2+}}$ for ${\rm Ca^{2+}}$ [44] %\cite{Zhang90}
does not offer a definitive test for the possibility of
negative charge cancellation by positive ions. This is because
high frequency sound absorption studies of
substitution rate constants for water molecules in the inner hydration
shell of various ions reveal that the inner hydration shell water
substitution rate for ${\rm Mg^{2+}}$ is more than three orders of
magnitude slower than that for ${\rm Ca^{2+}}$ [54], %\cite{Diebler69},
indicating that the positive charge of ${\rm Mg^{2+}}$ is shielded
much more effectively by water than is the case for ${\rm Ca^{2+}}$.
\newline \newline \noindent 
Thus, the slowing or stopping of anaphase-A
chromosome motion accompanying free
calcium concentration increases above the optimum concentration
for maximum anaphase-A chromosome speed -- but well below
concentration levels that compromise the mitotic apparatus --
is completely consistent with an electrostatic disassembly motor for
poleward chromosome motion.
This experimental observation has not been addressed by any of
the other current models for anaphase-A motion.
\section{Summary} 
It seems clear that cellular electrostatics involves more than the 
traditional thinking regarding counterion screening of electric fields and 
the resulting unimportance within cells 
of the second most powerful force in nature. The reality may be that the 
evidence suggests otherwise, and that the resulting enhanced electrostatic 
interactions are more robust and act over greater distances than 
previously thought. One aspect 
of this is the ability of microtubules to extend the reach of 
electrostatic force over cellular distances; another lies in the reduced 
counterion sceeening and dielectric constant of the cytosol between 
charged protein surfaces. 
\newline \newline \noindent
High ${\rm pH_i}$ during prophase favors spindle assembly. This includes 
greater electrostatic attractive forces between tubulin dimers as well as 
increased repulsive electrostatic interactions driving poleward movement 
of forming half-spindles. 
Additionally, because of significantly reduced counterion 
screening and the low 
dielectric constant of layered water adhering to charged free
ends of tubulin dimers, the necessary attraction and alignment of
tubulin during spindle self-assembly would be
enhanced by the considerably increased range and strength of the
electrostatic attraction between oppositely charged regions of
tubulin dimers. 
Similarly, the
mutually repulsive electrostatic force between a continually 
changing subset of 
like-charged plus ends of interacting microtubules from opposite
half-spindles in the growing mitotic spindle would be expected to be
significantly increased in magnitude and range.  
\newline \newline \noindent
Experimentally observed increases in whole cell
sialic acid content and intracellular pH during prophase, followed by an
observed release of free calcium from nuclear envelope and
endoplasmic reticulum stores, will
significantly enhance the expression of %%%MAGNITUDE OF
negative charge on sialic acid residues of
the nuclear envelope, providing sufficient electrostatic
energy for nuclear envelope breakdown.
Since terminal sialic acids are attached to membrane proteins that
are firmly anchored in the lipid bilayer,
the observed disassembly of the nuclear envelope into membrane 
fragments is consistent with 
electrostatic repulsion between membrane charge clusters that could 
tear apart under the influence of increased electrostatic charge.
\newline \newline \noindent
The observed lowering of both intracellular pH and whole cell sialic acid
content during late anaphase and telophase is consistent with a
decreased manifestation of net negative charge on 
membrane fragments. 
This decrease could shift the balance of thermal energy versus
electrostatic repulsive energy in favor of thermal energy,
allowing the closer approach of membrane fragments necessary for
reassembly biochemistry to occur in nascent daughter cells.
\newline \newline \noindent 
Changes in microtubule dynamics are integral to changes in the motions of 
chromosomes during mitosis. These changes in microtubule dynamics can be 
attributed to an associated 
change in intracellular pH (${\rm pH_i}$) during mitosis. In particular,  
a decrease in ${\rm pH_i}$ -- from a peak during prophase -- through 
mitosis may act as a master clock 
controlling microtubule disassembly/assembly probability ratios by 
altering the electrostatic interactions of tubulin dimers. 
This, in turn, could determine the timing and 
dynamics of post-attachment mitotic chromosome motions through metaphase. 
\newline \newline \noindent 
Force generation for the dynamics of post-attachment chromosome 
motions during prometaphase and metaphase can be explained 
by statistical fluctuations in nanoscale repulsive 
electrostatic microtubule antipoleward assembly forces acting between 
microtubules and chromosome arms in conjunction with similar 
fluctuations in nanoscale attractive electrostatic microtubule poleward 
disassembly 
forces acting at kinetochores and spindle poles [2,3]. The different 
motions throughout prometaphase and metaphase may be understood 
as an increase in the microtubule disassembly to assembly probability 
ratio due to a steadily decreasing ${\rm pH_i}$ [2,8]. 
\newline \newline \noindent 
Thus it seems reasonable to assume that the shift from
the dominance of microtubule growth during prophase, to a lesser 
extent during prometaphase, and to %%%%%%%virtual %%%OVERFULL>>>>>> 
approximate parity between microtubule polymerization 
and depolymerization during 
metaphase chromosome oscillations can be attributed to the 
gradual downward ${\rm pH_i}$ shift during mitosis %from a prophase peak 
that is observed in many cell types.
\newline \newline \noindent 
Evidence for a further continuing decrease in ${\rm pH_i}$ and an 
increasing microtubule disassembly to
assembly probability ratio is seen in increased
kinetochore tension just prior to anaphase.
This increased tension has a possible simple interpretation
in terms of the greater
magnitude of poleward electrostatic disassembly forces at 
kinetochores and poles relative to
antipoleward assembly forces between plus ends of microtubules and 
chromosome arms. %%%  AND POSSIBLY 
\newline \newline \noindent 
Additional continuing decreases in ${\rm pH_i}$ during 
anaphase-A and anaphase-B 
may not be the major determinant of anaphase motions [3,8,21]. 
\newline \newline \noindent
In light of the large body of experimental information regarding mitosis,
the complexity and lack of unity of models for the various 
events and  motions gives, at least to this observer, reason to believe 
that approaching mitosis primarily within the molecular biology 
paradigm is flawed. This paper reviews the merits of an approach 
based on continuum electrostatics. %%%%%%% AN APPROACH...%% %%%%%%%%%%%%%
%nanoscale electrostatic interactions. 
Such an approach to mitotic  
motions based on stably bound charge distributions can be used 
to frame a minimal assumptions model that incorporates the force 
production, timing, and sequencing of post-attachment chromosome motions. 
\section*{References} 
\small \mbox{} 
[1] L.J. Gagliardi, J. Electrostat. 54 (2002) 219. \newline \noindent
[2] L.J. Gagliardi, Phys. Rev. E 66 (2002) 011901. \newline \noindent
[3] L.J. Gagliardi, J. Electrostat. 63 (2005) 309. \newline \noindent  
%L.J. 2005. Electrostatic force generation in chromosome
%motions during mitosis. J. Electrostat. {\bf63}:309. \newline \newline   
%[4] E.L. Grishchuk {\it et al.}, 
%Proc. Natl. Acad. Sci. USA 105 (2008) 6918. \newline \noindent 
%E.L. 2008b. Different assemblies of the DAM1
%complex follow shortening microtubules by distinct mechanisms.
[4] C. Ciferri {\it et al.}, Cell 133 (2008) 427. \newline \noindent 
[5] G.J. Guimaraes, Y. Dong, B.F. McEwen, J.G. DeLuca,  
Current Biol. 18 (2008) 1778. \newline \noindent 
%G.J., Dong Y., McEwen B.F., DeLuca J.G. 2008.
%Kinetochore-microtubule attachment relies on the disordered N-terminal
%tail domain of Hec1. 
[6] S.A. Miller, M.L. Johnson, P.T. Stukenberg, Curr. Biol. 18 (2008) 
1785. \newline \noindent 
[7] S. Horme\~{n}o {\it et al.}, 
Biophys. J. 97 (2009) 1022.  \newline \noindent 
%2009. Single centrosome manipulation reveals
%its electric charge and associated dynamic structure. 
[8] L.J. Gagliardi, {\it Electrostatic Considerations in Mitosis}. 
iUniverse Publishing Co., Bloomington, IN, 2009. \newline \noindent 
[9] G.B.Benedek, F.M.H. Villars, {\it Physics: With
Illustrative Examples From Medicine and Biology: Electricity and
Magnetism}. Springer-Verlag, 2000, p. 403. \newline \noindent
[10] B. Alberts, D. Bray, J. Lewis, M. Raff, M.K. Roberts,
J.D. Watson. {\it Molecular Biology of the Cell}. 
Garland Publishing Co., N.Y., 1994, p. 920. \newline \noindent 
[11] C. Amirand {\it et al.}, 
Biol. Cell 92 (2000) 409. \newline \noindent 
%2000. Intracellular pH in one-cell mouse embryo
%differs between subcellular compartments and between interphase and
%mitosis. 
[12] R.A. Steinhardt, M. Morisawa, In: R. Nuccitelli,
D.W. Deamer (Eds.), Intracellular pH:  Its Measurement, Regulation, and
Utilization in Cellular Functions. Alan R. Liss, New York, 1982, pp.
361-374. \newline \noindent
[13] G. Schatten, T. Bestor, R. Balczon, 
Eur. J. Cell Biol. 36 (1985) 116. \newline \noindent 
%1985. Intracellular pH shift
%leads to microtubule assembly and microtubule-mediated motility
%during sea urchin fertilization:  Correlations between elevated
%intracellular pH, microtubule activity and depressed
%intracellular pH and microtubule disassembly. 
[14] M.W. Kirschner, J. Cell Biol. 86 (1980) 330. \newline \noindent 
%1980. Implications of treadmilling for the stability and
%polarity of actin and tubulin polymers in vivo. \newline \noindent 
[15] M. De Brabander, G. Geuens, R. Nuydens, 
Cold Spring Harbor Symp. Quant. Biol. 46 (1982) 227. \newline \noindent 
%1982. Microtubule stability and
%assembly in living cells:  The influence of metabolic inhibitors, taxol 
%and pH. 
[16] W.J. Deery, B.R. Brinkley, 
J. Cell Biol. {\bf96}:1631.\newline \noindent 
%1983. Cytoplasmic microtubule-disassembly from
%endogenous tubulin in a Brij-lysed cell model. 
[17] B. Alberts, D. Bray, J.Lewis, M. Raff, M.K. Roberts, 
J.D. Watson,  {\it Molecular Biology of the Cell}. New
York: Garland Publishing Company, 1994, p. 1041. \newline \noindent
[18] D. Jordan-Lloyd, A. Shore, {\it The Chemistry of Proteins}. 
London: J. A. Churchill Publishing Company, 1938  \newline \noindent          
[19] L. Pauling, J. Am. Chem. Soc. 67 (1945) 555. \newline \noindent
%1945. The adsorption of water by proteins. 
[20] M.F. Toney, J.N. Howard, J. Richer, G.L. Borges, J.G. Gordon, 
O.R. Melroy, D.G. Wiesler, D. Yee, L. Sorensen, 
Nature 368 (1994) 444. \newline \noindent 
%1994. Voltage-dependent ordering of
%water molecules at an electrode-electrolyte interface. 
[21] L.J. Gagliardi, J. Electrostat. 66 (2008) 147. \newline \noindent
[22] H.C. Joshi, M.J. Palacios, L. McNamara, D.W. Cleveland, 
Nature 356 (1992) 80. \newline \noindent 
%1992. $\gamma$-Tubulin is a centrosomal protein required for cell
%cycle-dependent microtubule nucleation. 
[23] S.L. Wolfe, 
{\it Molecular and Cellular Biology}. Belmont, CA: Wadsworth
Publishing Company, p. 1012. \newline \noindent 
%1993a. 
[24] R. Heald, R. Tournebize, T. Blank, R. Sandaltzopoulos, P. Becker, 
A. Hyman, E. Karsenti, 
Nature 382 (1996) 420. \newline \noindent  
%1996. Self-organization of microtubules into bipolar
%spindles around artificial chromosomes in Xenopus egg extracts. 
[25] M.V. Satari\'{c}, J.A. Tuszy\'{n}ski, R.B.
\u{Z}akula, Phys. Rev. E, 48 (1993) 589. \newline \noindent
[26] J.A. Brown, J.A. Tuszy\'{n}ski, Phys. Rev. E 56 (1997) 5834. 
\newline \noindent
[27] N.A. Baker, D. Sept, S. Joseph, M.J. Holst, J.A.
McCammon, Proc. Nat. Acad. Sci. 98 (2001) 10037.
\newline \noindent
[28] J.A. Tuszy\'{n}ski, J.A. Brown, P. Hawrylak, Phil. Trans. 
R. Soc. Lond. A 356 (1998) 1897. \newline \noindent
[29] J.A. Tuszy\'{n}ski, S. Hameroff, M.V. Satari\'{c}, B.
Trpisov\'{a}, M.L.A. Nip, J. Theor. Biol. 174 (1995) 371. \newline
\noindent
[30] R. Stracke, K.J. B\"{o}hm, L. Wollweber, J.A.
Tuszy\'{n}ski, E. Unger, Bioch. and Biophys. Res. Comm. 293 (2002) 602.
\newline \noindent
[31] D. Sackett, pH-induced conformational changes in the
carboxy-terminal tails of tubulin, presented at the Banff Workshop
Molecular Biophysics of the Cytoskeleton, Banff, Alberta, Canada, August
25-30, 1997. \newline \noindent
[32] J.A. Tuszy\'{n}ski, J.A. Brown, E.J. Carpenter, E. Crawford. 2002. 
In: Proceedings of the Electrostatics Society of America and Institute 
of Electrostatics - Japan. Morgan Hill, CA: Laplacian Press, pp. 41-50. 
\newline \noindent
%[33] G. Schatten, G.G. Maul, H. Schatten, European J. Cell
%Biol. 36 (1985) 116. \newline \noindent
[33] R.C. Weisenberg, 
Science 177 (1972) 1104. \newline \noindent 
%1972. Microtubule formation {\it in vitro} in solutions
%containing low calcium concentrations. 
[34] G.G. Borisy, J.B. Olmsted, 
Science 177 (1972) 1196. \newline \noindent 
%1972. Nucleated assembly of microtubules
%in porcine brain extracts. 
[35] G.S. Hartley, J.W. Roe, 
Trans. Faraday Soc. 35 (1940) 101. \newline \noindent 
[36] L.J. Gagliardi, J. Electrostat. 64 (2006) 843. \newline \noindent 
[37] M.C. Glick, E.W. Gerner, and L. Warren, J. Cell 
Physiol. 77 (1971) 1. \newline \noindent 
[38] R.B. Silver, L.A. King, and A.F. Wise, Biol. Bull. 
195 (1998) 209. \newline \noindent 
[39] R.B. Silver, Biol. Bull. 187 (1994) 235. \newline \noindent 
[40] L. Weiss, J. Theoret. Biol. 18 (1968) 9. \newline \noindent 
[41] A. Grancell, P.K. Sorger, 
Current Biol. 8 (1998) R382. \newline \noindent 
%1998. Chromosome movement:  Kinetochores motor along. 
[42] H. Maiato, J. DeLuca, E.D. Salmon, W.C. Earnshaw, 
J. Cell Science 117 (2004) 5461. \newline \noindent
%2004. The dynamic kinetochore-microtubule interface. 
%1940. Ionic concentrations at interfaces. 
[43] Y.H. Song, E. Mandelkow, J. Cell Biol. 128 (1995) 81. 
\newline \noindent 
[44] D.H. Zhang, D.A. Callaham, P.K. Hepler, J. Cell Biol. 111 (1990) 
171. \newline \noindent  
%%[45] R.C. Weisenberg, Science 177 (1972) 1104. \newline \noindent 
[45] E.D. Salmon, R.R. Segall, J. Cell Biol. 86 (1980) 355. 
\newline \noindent 
[46] D.P. Kiehart, J. Cell Biol. 88 (1981) 604. \newline \noindent 
[47] W.Z. Cande, Physiology of chromosome movement in 
lysed cell models, In: H.G. Schweiger (Ed.), International Cell 
Biology, Springer, Berlin, 1981, pp. 382-391. \newline \noindent 
[48] J.B. Olmsted, G.G. Borisy, Biochemistry 14 (1975) 2996. 
\newline \noindent 
[49] R.B. Nicklas, Chromosome movement: current models and
experiments on living cells, In: S. Inoue, R.E. Stephens (Eds.), Molecules
and Cell Movement, Raven Press, New York, 1975, pp. 97-117. 
\newline \noindent 
[50] R.B. Nicklas, Chromosomes and kinetochores do more in
mitosis than previously thought, In: J.P. Gustafson, R. Appels, R.J.
Kaufman (Eds.), Chromosome Structure and Function: The Impact of New 
Concepts, Plenum, New York, 1987, pp. 53-74. \newline \noindent 
[51] E.D. Salmon, Ann. N.Y. Acad. Sci. 253 (1975) 383. 
\newline \noindent 
[52] E.D. Salmon, Microtubule dynamics and chromosome
movement, In: J.S. Hyams, B.R. Brinkley (Eds.), Mitosis: Molecules
and Mechanisms, Academic Press, San Diego, 1989, pp. 119-181. 
\newline \noindent 
[53] S.L. Wolfe, Molecular and Cellular Biology, second ed., 
Wadsworth, Belmont, CA., 1993, p. 425. \newline \noindent 
[54] H. Diebler, G. Eigen, G. Ilgenfritz, G. Maass, R.
Winkler, Pure Appl. Chem. 20 (1969) 93.

\end{document}